\documentclass[sigconf]{acmart}

\usepackage{array}
\usepackage[normalem]{ulem}

\AtBeginDocument{%
  }

\copyrightyear{2026}
\acmYear{2026}
\setcopyright{cc}
\setcctype{by}
\acmConference[CHI '26]{Proceedings of the 2026 CHI Conference on Human Factors in Computing Systems}{April 13--17, 2026}{Barcelona, Spain}
\acmBooktitle{Proceedings of the 2026 CHI Conference on Human Factors in Computing Systems (CHI '26), April 13--17, 2026, Barcelona, Spain}
\acmDOI{10.1145/3772318.3790395}
\acmISBN{979-8-4007-2278-3/2026/04}

\begin{document}

\title[Convivial Fabrication]{Convivial Fabrication: Towards Relational Computational Tools For and From Craft Practices}

\author{Ritik Batra}
\affiliation{%
  \department{Information Science}
  \institution{Cornell Tech}
  \city{New York, New York}
  \country{USA}
}
\email{rb887@cornell.edu}

\author{Roy Zunder}
\affiliation{%
  \institution{Cornell Tech}
  \city{New York, New York}
  \country{USA}}
\email{royzun@gmail.com}

\author{Amy Cheatle}
\affiliation{
  \department{Science and Technology Studies}
  \institution{Cornell University}
  \city{Ithaca, New York}
  \country{USA}}
\email{ac2288@cornell.edu}

\author{Amritansh Kwatra}
\affiliation{
  \department{Information Science}
  \institution{Cornell Tech}
  \city{New York, New York}
  \country{USA}}
\email{ak2244@cornell.edu}

\author{Ilan Mandel}
\affiliation{
  \department{Information Science}
  \institution{Cornell Tech}
  \city{New York, New York}
  \country{USA}}
\email{im334@cornell.edu}

\author{Thijs Roumen}
\affiliation{
  \department{Information Science}
  \institution{Cornell Tech}
  \city{New York, New York}
  \country{USA}}
\email{thijs.roumen@cornell.edu}

\author{Steven J. Jackson}
\affiliation{
  \department{Information Science}
  \institution{Cornell University}
  \city{Ithaca, New York}
  \country{USA}}
\email{sjj54@cornell.edu}

\renewcommand{\shortauthors}{Batra et al.}

\begin{abstract}
Computational tools for fabrication often treat materials as passive rather than active participants in design, abstracting away relationships between craftspeople and materials. For craft communities that value relational practices, abstractions limit the adoption and creative uptake of computational tools which might otherwise be beneficial. To understand how better tool design could support richer relations between individuals, tools, and materials, we interviewed expert woodworkers, fiber artists, and metalworkers. We identify three orders of convivial relations central to craft: immediate relations between individuals, tools, and materials; mid-range relations between communities, platforms, and shared materials; and extended relations between institutions, infrastructures, and ecologies. Our analysis shows how craftspeople engage and struggle with convivial relations across all three orders, creating workflows that learn from materials while supporting autonomy. We conclude with design principles for computational tools and infrastructures to better support material dialogue, collective knowledge, and accountability, along with richer and more convivial relations between craftspeople, tools, and the material worlds around them.

\end{abstract}

\begin{CCSXML}
<ccs2012>
   <concept>
       <concept_id>10003120.10003121.10011748</concept_id>
       <concept_desc>Human-centered computing~Empirical studies in HCI</concept_desc>
       <concept_significance>500</concept_significance>
       </concept>
   <concept>
       <concept_id>10003120.10003121.10003126</concept_id>
       <concept_desc>Human-centered computing~HCI theory, concepts and models</concept_desc>
       <concept_significance>500</concept_significance>
       </concept>
 </ccs2012>
\end{CCSXML}

\ccsdesc[500]{Human-centered computing~Empirical studies in HCI}
\ccsdesc[500]{Human-centered computing~HCI theory, concepts and models}

\keywords{Craft, Conviviality, Fabrication, Computing}

% \received{11 September 2025}
% \received[revised]{4 December 2025}
% \received[accepted]{15 January 2026}

\maketitle

\section{Introduction}

Computational tools for fabrication have made important contributions to craft practice by advancing precision, repeatability, and design iteration.
Tools such as Computer-Aided Design (CAD) programs support craftspeople in translating their ideas into digital representations, such as vector paths or 3D models.
Digital abstractions facilitate the generation of plans for machine-enabled fabrication~\cite{gershenfeld_fab_2005}, serve as references for handcrafts, and assist with knowledge sharing within communities~\cite{torrey_learning_2009}.
Craftspeople use these tools to explore and iterate on designs before committing to physical workflows~\cite{hirsch_nothing_2023}. 
In other cases, computational tools enable form and processes previously unavailable---as in fine arts furniture maker Wendell Castle's stack lamination technique~\cite{cheatle_digital_2015}.
Automation and low-cost replication have produced important, if partial, gains for some craft practices and communities. 
These tools reshape craft toward speed, efficiency, new formal possibilities, and predictable execution regardless of material variability.

However, computational tools also sometimes interfere with a craftsperson's relationship with their materials and the worlds from which they come~\cite{mccullough1998abstracting}. 
Craft frequently demands intimacy with the temper, texture, and grain of a material. However, tools for computational fabrication often embed what anthropologist Tim Ingold has criticized as a \emph{hylomorphic} model of making and design, characterized by \emph{``the imposition of preconceived form upon a formless and inert substance''}~\cite{ingold_making_2013}.
Constrained by these tools, craft becomes `artifact-centric' by rigidly following predetermined plans to produce a final object.
Hylomorphic models leave no space for materiality to insert itself and reshape the making process.
Craftspeople may become estranged from the origins of their materials, losing embodied knowledge and accountability while reproducing the fallacy of `raw' materials drawn from a flat and uniform `standing reserve'~\cite{heidegger1977question}.
Computational tools may contribute to patterns of automation and deskilling---both within systems of mass production and in sectors previously excluded from such practices---that may challenge worker autonomy, organizing, and justice~\cite{nobleforces, zuboff1988age, khovanskaya2019tools}. 
In all these ways, computational tools have challenged craft-based modes of fabrication, reducing the capacity of forces outside the designer to `push back' and undermining opportunities for discovery, surprise, and co-creation long central to craft practices.

Computational tools for fabrication often fail to account for this co-creation.
While laser cutting wood, the tool-path is defined in illustration software by a vector, remaining ignorant of whether the laser will struggle to burn through a burl or follow the grain~\cite{baudisch_kyub_2019}.
CAD programs define models for 3D printing with mathematical precision, offering no language for the way materials shrink, warp, or slump at different temperatures~\cite{mcelroy_potscript_2023}.
Knitting software creates fixed patterns that do not adapt to the elasticity, twist, or slubs of a yarn as it runs through the feeder~\cite{hofmann_knitscript_2023}.
Our tools demand predictable execution of a predetermined plan rather than open-ended engagement with material realities.
As a result, our making processes may be wasteful and out of step with craft norms and practices~\cite{devendorf_probing_2016}.
Even when we provide `material aware' tools or sensors---for example, \emph{Sensicut}~\cite{dogan_sensicut_2021} or \emph{XR-Penter}~\cite{iyer_xr-penter_2025}---they still operate within a hylomorphic model in which variability is a problem to be corrected rather than a quality to be engaged.
Fabrication becomes the execution of an optimized plan, leaving little room for the craftsperson’s responsiveness, improvisation, or judgment.

In this work, we explore alternatives to the hylomorphic model through fabrication workflows that foreground craftsperson autonomy and responsive dialogue with materials. 
To ground these alternatives, we turn to Ivan Illich's distinction between \textit{convivial} tools (tools that enhance a person's creativity and autonomy) and \textit{non-convivial} tools (tools that create dependency and narrow engagement with the material world)~\cite{illich_tools_1973}.
A bicycle is convivial because it amplifies human movement while remaining under the rider's control.
An automobile, by contrast, is non-convivial because it is isolating and limited by highway infrastructures.
Convivial tools invite creative use and adaptation, while non-convivial tools demand compliance.
Convivial tools are widely distributed and can be used by anyone, while non-convivial tools remain limited by the `managers' controlling them.
Lastly, convivial tools deepen relations and accountability to the worlds around them, while non-convivial tools may encourage shallower relations of use and consumption.

Drawing on 23 interviews with craftspeople (woodworkers, fiber artists, and metalworkers), we identify three distinct yet interconnected orders of convivial relations that operate in craft practices:

\begin{enumerate}
    \item the immediate relation with hands-on dialogue between the craftsperson, tool, and material;
    \item the mid-range relation, where material knowledge is shared with communities through platforms; and
    \item the extended relation, where infrastructures connect institutions to their broader environmental and cultural ecologies.
\end{enumerate}

\textbf{In this paper, we argue that extending Illich's concept of conviviality to include these three orders of relations can guide the HCI community towards tools that support (rather than abstract) human-material collaboration in craft workflows while going beyond the traditional artifact-centrality of the field.}
From these orders, we offer seven design principles for computational tools that support convivial relations: embrace ``imperfect'' materials, develop ``fuzzy'' models for fabrication, synchronize fabrication pacing with natural rhythms, support evolving collective knowledge, bridge knowledge across orders, implement fault tolerance, and prioritize modularity with shared data.
We explore how conviviality can be demonstrated within and across the orders, offering a broader range of pathways for designing technologies that support craftspeople’s autonomy and meaningful relationships with materials, communities, and institutions.

We begin with a review of related work, positioning our work at the intersection of HCI, anthropology, and philosophy.
We then introduce our methodology for analyzing interviews with craftspeople and present our findings through the lens of our three orders of convivial relations.
Finally, we discuss how this expanded understanding of conviviality offers new analytical and design principles that extend beyond fabrication to foreground autonomy, personal expression, and responsive dialogue with materials.

\section{Related Work}

Our work exploring computing, craft, and materials draws on three branches of related work: materiality and fabrication in HCI, morphogenetic models of making, and convivial tools in computing.

\subsection{Materiality and Fabrication in HCI}

Research in HCI, computational design, and fabrication has embraced digitizing physical materials for computational manipulation.
Neil Gershenfeld's vision for \emph{``personal fabricators [to] bring the programmability of the digital world to the rest of the world''}~\cite{gershenfeld_fab_2005} exemplifies this approach, positioning materials as programmable substrates that can be modeled and simulated using computing.
Patrick Baudisch similarly envisions digital and physical objects as interchangeable: \emph{``(1) The scanner is a hardware unit that turns physical objects into digital objects, an 'analog-to-digital converter' (AD). (2) The printer is a hardware unit that turns virtual objects into physical objects, a 'digital-to-analog converter' (DA)''}~\cite{baudisch_personal_2017}.
Hiroshi Ishii argues for a future where physical materials \emph{``can change form and appearance dynamically, so they are as reconfigurable as pixels on a screen''}, further illustrating how computational thinking structures, and therefore constrains our understanding of materials~\cite{ishii_radical_2012}.

The HCI community has developed sophisticated methods of digitally representing artifacts to minimize variability and uncertainty in fabrication workflows.
Digital fabrication research has proposed automatic assembly systems~\cite{roumen_autoassembler_2021}, joint design systems to prevent assembly errors~\cite{park_foolproofjoint_2022}, and optimizations to fix structural issues in digital models~\cite{abdullah_fastforce_2021}.
Tools such as \emph{faBrickation} substitute fabrication material with standardized building blocks to optimize for speed~\cite{mueller_fabrickation_2014}, while systems to 3D print with yarn apply abstractions designed for plastic filament to fiber~\cite{hudson_printing_2014}.
These abstractions build on a long history of computer graphics research to visualize materials using computational modeling~\cite{pharr_physically_2023, tewari2020state} that extend to industry-grade systems such as Autodesk Fusion where designers model material-agnostic geometries and only at the end of their design process incorporate materials as colors and textures for rendering.\footnote{\href{https://web.archive.org/web/20260127133547/https://forums.autodesk.com/t5/fusion-design-validate-document/method-for-creating-wood-material-in-fusion/td-p/5459477}{https://forums.autodesk.com/t5/fusion-design-validate-document/method-for-creating-wood-material-in-fusion/td-p/5459477}}
The field is dominated by computational tools that prioritize predictable outcomes for fabrication by eliding materiality.

While such approaches have made foundational contributions to HCI work in accessible, portable, and sustainable fabrication, they prioritize fabrication speed, reproducibility, and novice usability in ways that may limit their utility and adoption in craft contexts~\cite{adamson_thinking_2018}.
The systematic separation between design in digital space and making in material space~\cite{twigg-smith_tools_2021} creates homogeneity that recognizes certain material engagements while rendering others invisible~\cite{star_ethnography_1999}.
Even where tools are explicitly aimed toward adoption within craft contexts---for example, \emph{Spyn}~\cite{rosner2008spyn}, \emph{MatchSticks}~\cite{tian_matchsticks_2018}, and \emph{FreeD}~\cite{zoran_freed_2013}---these \emph{hybrid} tools~\cite{liu2024learning, bonanni2008future} still constrain the material responsiveness, community relationships, and ecological considerations that typically drive craft practices~\cite{maceachren2000crafting}.
In the following sections, we argue for two important but missing relations of computational tools for fabrication: \emph{morphogenetic} design with materials and \emph{convivial} relations with extended worlds.

\subsection{Morphogenetic Models of Making}

Digital fabrication's paradigm of imposing digital form onto passive matter can be characterized by Ingold's description of a \emph{hylomorphic} model of making.
As an alternative, Ingold proposes a \emph{morphogenetic} model, where the maker's role is to \emph{``follow the materials''} by engaging in \emph{``an ongoing generative movement that is at once itinerant, improvisatory and rhythmic''}~\cite{ingold_textility_2010}.
The way a beaver constructs a dam is exemplary of this method; as Ingold observes, \emph{``the beaver... inhabits an environment that has been decisively modified by the labours of its forbears, in building dams and lodges, and will in turn contribute to the fashioning of an environment for its progeny''}~\cite{ingold_perception_2021}.

An exemplar of the consequences of hylomorphic thinking is evident in Santiago Calatrava's architectural pieces and his \emph{``stubborn devotion to form''}~\cite{daley_star_2013}.
Although admired for his sculpture-like buildings, Calatrava's designs are critiqued for being often over budget, delayed, and defective by prioritizing the form over material constraints.
The Queen Sofía Palace of the Arts in Valencia designed by Calatrava, for example, features a metal shell clad in mosaic tiles (trencadís) but failed to account for how metal and ceramic expand at different rates under daily temperature swings.
The building became a safety hazard as tiles warped and fell off. Repairs for this warping included removing 8,000 square meters of tiles (costing 3 million euros) and painting the exposed metal white as a temporary solution~\cite{minder_spanish_2014}.
This one of many examples of the consequences of ignoring the material rather than \emph{``corresponding to''} them as collaborators with their own properties, limits, and behaviors over time~\cite{ingold_correspondences_2021}.

HCI research is increasingly exploring how computational tools can follow Ingold's morphogenetic model of making~\cite{ingold_textility_2010}.
\citet{devendorf_probing_2016} challenges traditional fabrication approaches by creating \emph{Redeform}, a system where humans follow machine-generated paths using materials like flowers and charcoal, allowing material resistance to actively shape fabrication outcomes.
Projects such as \emph{CoilCAM}~\cite{bourgault_coilcam_2023} and \emph{Throwing Out Conventions}~\cite{moyer_throwing_2024} further demonstrate how computational tools can respond to material behaviors.
Similarly, \citet{twigg-smith_tools_2021} analyze \emph{\#PlotterTwitter} to reveal how maker communities develop workflows that \emph{``champion creative exploration of interwoven digital and physical materials over predictable fabrication steps.''}
In a more literal sense, research focused on bio-based materials such as shape-changing clay-dough~\cite{bell_shape-changing_2024}, composted food waste~\cite{bell_reclaym_2022}, and living materials such as bacterial cellulose~\cite{nicolae_biohybrid_2023} further embrace the role of materials in what they describe as \emph{``material-led design''}~\cite{benabdallah_chaotic_2025}. 

Ron Wakkary's \emph{``more-than-human''} approach~\cite{wakkary_things_2021} extends Ingold's arguments by critiquing the anthropocentric assumptions that often underlie HCI research. He notes instead how \emph{``humans share center stage with non-humans, and [are] bound together materially, ethically, and existentially''}~\cite{wakkary_things_2021}.
\citet{benabdallah_technical_2024} similarly critique the hylomorphic schema in digital fabrication, arguing for what Gilbert Simondon calls a \emph{``technical mentality''} which recognizes how materials and machines co-produce knowledge through their interactions.
This acknowledgment of non-human agency in \emph{``designing with''} the material world both aligns with and extends Ingold's morphogenetic model by considering not just materials but also the broader ecological contexts and forces within which materials exist~\cite{star_ethnography_1999}.

\subsection{Convivial Tools in Computing} 
Hylomorphic models of making are only one way of losing our relations to the pluriverse of worlds around us~\cite{escobar2018designs}. In 1973, Ivan Illich argued that the forces of progress and increasing specialization attached to modern industrial systems were leading to a loss of autonomy and a deep alienation from everyday life~\cite[p.103]{illich_tools_1973}.
The compulsive pursuit of growth, Illich observes, \emph{``establishes a radical monopoly not only over resources and tools but also over the imagination and motivational structure of people''}~\cite[p.103]{illich_tools_1973}.
The prognosis for uncontrolled expansion, he suggests, is disastrous: over-efficient tools destroy the balance between humans and nature while \emph{``upsett[ing] the relationship between what people need to do by themselves and what they need to obtain ready-made''}~\cite[p.62]{illich_tools_1973}.

The remedy, Illich argues, is a \emph{convivial society} built around the use of \textit{convivial tools}: tools designed as instruments that deepen human autonomy while recognizing and reconnecting our relations and responsibilities to the environments and worlds around us.
By tools, he means not only hardware (e.g., drills, hammers, cars), but also the broader institutions and systems that produce both material goods and intangible services such as education, communication, and medical care~\cite[p.34]{illich_tools_1973}.
Illich recognizes that such tools are always entangled within complex networks of social relationships and individual agency.
But without those values embedded into the design of the tool, it inevitably shapes the individual around its own logic and constraints (rather than the other way around): \emph{``To the degree that he masters his tools, he can invest the world with his meaning; to the degree that he is mastered by his tools, the shape of the tool determines his own self-image''}~\cite[p.29]{illich_tools_1973}.
The question of who (tool or individual) shapes whom becomes central to Illich's concept of conviviality, examining how tools either enhance or diminish human autonomy and creativity.

For Illich, convivial tools exist on the opposite end of the spectrum from industrial tools.
While industrial tools are manipulated to function within limits set by \emph{``the managers of our major tools-nations, corporations, parties, structured movements, professions-hold power, convivial tools [expand] the range of each person's competence, control, and initiative, limited only by other individuals' claims to an equal range of power and freedom''}~\cite[p.12]{illich_tools_1973}.
Hand tools such as hammers or pocket knives are typically convivial as they are \emph{``mere transducers of the energy generated by man’s extremities and fed by the intake of air and of nourishment''}~\cite[p.29]{illich_tools_1973}.
In contrast, externally powered tools such as jet planes and air conditioners \emph{``favor centralization of control''}~\cite[p.52]{illich_tools_1973}, leaving their users as operators rather than collaborators.
David Noble's record of computer numerical control (CNC) in manufacturing demonstrates how computational tools can embody industrial values as CNC machines were designed to shift control from skilled machinists to management, replacing craft knowledge with centralized control~\cite{nobleforces}.

Although influential in the social and environmental sciences, Illich's ideas around conviviality have been taken up only occasionally in HCI and related fields to date.
\citet{janaszek_convivial_2010}, for example, has described open-source and free software as convivial computational tools that encourage user freedom and inter-connectedness in contrast to closed-source software (such as Microsoft Office) that requires end user licenses and limits social engagements to those tools.
Other researchers have drawn on Illich to examine the relationship between internet social networks and social change~\cite{ameripour_conviviality_2010}, the role of artificial intelligence as a convivial tool for the marketing domain~\cite{rindfleisch_ai_2024}, sustainability and social justice in relation to computing~\cite{de2021pluriverse}, and the development of open-sourced agricultural technology~\cite{pantazis2020tools}.
HCI scholars have drawn on Illich to study the development of music software for interdependent communities of visually impaired and blind sound creatives~\cite{lucas_qualities_2025}, navigation of health information on social media in urban and rural South India~\cite{karusala_towards_2022}, and how assistive prototypes can empower rather than constrain individuals with disabilities~\cite{choueiri_can_2018}.

Most explorations of conviviality within HCI focus on digital interfaces and information systems, leaving underexplored how convivial principles might apply to computational tools that mediate material practices such as craft.
Although Illich does not explicitly use the language of craft, he invokes its values of autonomy and care when discussing handmade versus manufactured objects: \emph{``People need not only to obtain things, they need above all the freedom to make things among which they can live, to give shape to them according to their own tastes, and to put them to use in caring for and about others''}~\cite[p.20]{illich_tools_1973}.
Our work addresses this gap by examining how conviviality operates across the expanded relations that characterize craft practices, moving beyond individual tool use to consider the broader infrastructures, institutions, and ecologies that impact material engagements.
In the following section, we describe our work with expert craftspeople to understand how they experience, negotiate, and develop the relationships between their tools and materials.

\section{Methods}

\begin{table*}[ht]
\centering
\small
\begin{tabular}{
    >{\raggedright\arraybackslash}p{.06\columnwidth}
    >{\raggedright\arraybackslash}p{.04\columnwidth}
    >{\raggedright\arraybackslash}p{.5\columnwidth}
    >{\raggedright\arraybackslash}p{.7\columnwidth}
    >{\raggedright\arraybackslash}p{.6\columnwidth}}
    \toprule
    \textbf{ID} & \textbf{Exp.} & \textbf{Profession(s)} & \textbf{Techniques} & \textbf{Materials} \\
    \midrule
    P1 & 50+ & Sculptor, Book Artist, Fiber Artist, Teacher & Weaving, Knitting, Crochet, Sewing & Fabric, Yarn (Cotton, Acrylic) \\
    P2 & 20 & Studio Owner, Architect, Product Designer & CNC, 3D Printing, Machining, Sewing, Laser Cutting & Metal, Wood, Fabric, Stone \\
    P3 & 15 & Product Designer & 3D Printing, Knitting, Embroidery & Wood, Fabric, Cellulose Sponge \\
    P4 & 15+ & Sculptural Artist, Visual Arts Professor & Kinetic Sculptures, Knitting, Crochet, Digital Fabrication, Welding & Wood, Metal, Sugar, Plaster, Yarn \\
    P5 & 15+ & Furniture Designer & Marquetry, Weaving, CNC, Laser Cutting & Corn Husks, Agave Fibers, Loofah, Wood \\
    P6 & 10+ & Woodworker, Furniture Designer, Visual Arts Professor & Woodworking, Hand Carving, 3D Printing & Wood (Walnut, Oak, Cypress) \\
    P7 & 25+ & Sculptural Artist & Welding, Sculpture, Knitting, Laser Cutting & Plywood, Cardboard, Sheet Metal \\
    P8 & 20+ & Knitting Studio Manager & Knitting, Crochet & Yarn (Natural, Chenille) \\
    P9 & 15 & Jewelry Designer, Design Professor & Jewelry Making, Metalwork, Enameling, 3D Printing & Metal (Brass, Copper, Silver, Gold) \\
    P10 & 20 & Sculptural Artist, Visual Arts Professor & Sculpture, 3D Animation, Laser Cutting & Metal (Corten Steel), Wood \\
    P11 & 50+ & Textile Designer & Loom Weaving, Machine Knitting & Yarn (Natural, Synthetic) \\
    P12 & 25 & Metalworker, Makerspace Director & Welding, Metal Fabrication, CNC, Laser Cutting & Metal (Steel) \\
    P13 & 20 & Weaving Artist, Studio Manager & Floor Loom Weaving & Yarn (Cotton, Wool, Linen, Nylon) \\
    P14 & 25+ & Hobbyist, Student & Knitting, Crochet, Pottery, 3D Printing & Yarn (Acrylic), Clay, Straw, Flax \\
    P15 & 5+ & Hobbyist, Small Engine Repair, Software Engineer & Leatherwork, Metalwork, Welding, Laser Cutting & Leather, Metal (Steel) \\
    P16 & 10+ & Community Organizer, Fiber Artist & Crochet, Knitting, Machine Knitting, Weaving & Yarn (Acrylic, Cotton, Wool, Silk) \\
    P17 & 10+ & Woodworker, Teacher & Green Woodworking, Chair Making, Hand Tools & Wood (Oak, Maple) \\
    P18 & 30+ & Mathematician, Sculptural Artist, Math Lecturer & Crochet, Basket Weaving & Yarn (Polyester, Acrylic), Wire \\
    P19 & 35 & Sculptural Artist, Community Garden Coordinator & Sculpture, Mosaics, Welding & Metal (Steel), Glass, Wood (Black Locust), Mycelium, Bamboo \\
    P20 & 35 & Woodworker, Software Engineer & Woodworking, Hand Tools, Model Ship Building & Wood (Cherry, Oak, Pine), Plywood \\
    P21 & 5+ & Furniture Maker, Teacher, Drafter & Woodworking, Hand Drafting, CAD & Wood (Hardwoods, Softwoods) \\
    P22 & 5+ & Mending Artist, Teacher & Embroidery, Mending, Stitching & Fabric (Upcycled, Linens, Polyester) \\
    P23 & 15 & Textile Artist, Woodworker, Freelancer & Print Media, Woodworking, Ceramics, Sculpture, Sewing, Tufting & Fabric (Upcycled), Wood, Clay, Vinyl \\
    \bottomrule
\end{tabular}
\caption{Summary of expert participants (P), their years of professional experience, professions, primary craft techniques, and materials discussed.}
\Description{This table outlines all the participants in our expert craft interview study. It includes their years of experience as well as the techniques and materials they shared during the interviews.}
\label{tab:participants}
\end{table*}

In this work, we conduct and qualitatively analyze 23 semi-structured interviews between July 2024 and April 2025.
Interviews were conducted both in-person and over Zoom, ranging in length from 30 to 75 minutes.
Our inquiry was initially guided by the question: \emph{How do craftspeople negotiate the tensions between materials and computing to achieve sustainable workflows?}
The interview questions focused on the participants' professional backgrounds, use of tools in their practice, relationships with the materials, and perspectives on environmental sustainability (see Supplementary A1 for the full interview protocol).

\subsection{Craft as a Site of Inquiry}

We understand craft practices as rich sites of inquiry because as \citet{frankjaer_understanding_2018} argue, it is not only a set of fabrication techniques but a mode of making that extends beyond material boundaries into computational domains and logics.
Craft is characterized by \citet{csikszentmihalyi_meaning_1981} as a fabrication process where craftspeople remain responsible for all elements of production, either directly or via close collaboration with others. 
HCI researchers have frequently turned to examinations of craft to better understand the negotiated space where computation intersects with deeply embodied and traditional practices \cite{cheatle_recollecting_2023, jacobs_digital_2016, rosner_reflections_2009}.

Craftspeople must decide when to foreground computational tools across diverse contexts.
Such decisions demand deep knowledge of both computational systems and material affordances~\cite{frankjaer_understanding_2018}.
Studying craft practices provides a unique opportunity to examine the (1) tacit processes through which craftspeople transition between computational and hand tools, (2) the tensions involved in such transitions, and (3) the forms of embedded knowledge that emerge from reconciling these tensions~\cite{sennett_craftsman_2008}.

\subsection{Participants and Recruitment}
\begin{figure*}[ht]
    \centering
    \includegraphics[width=\textwidth]{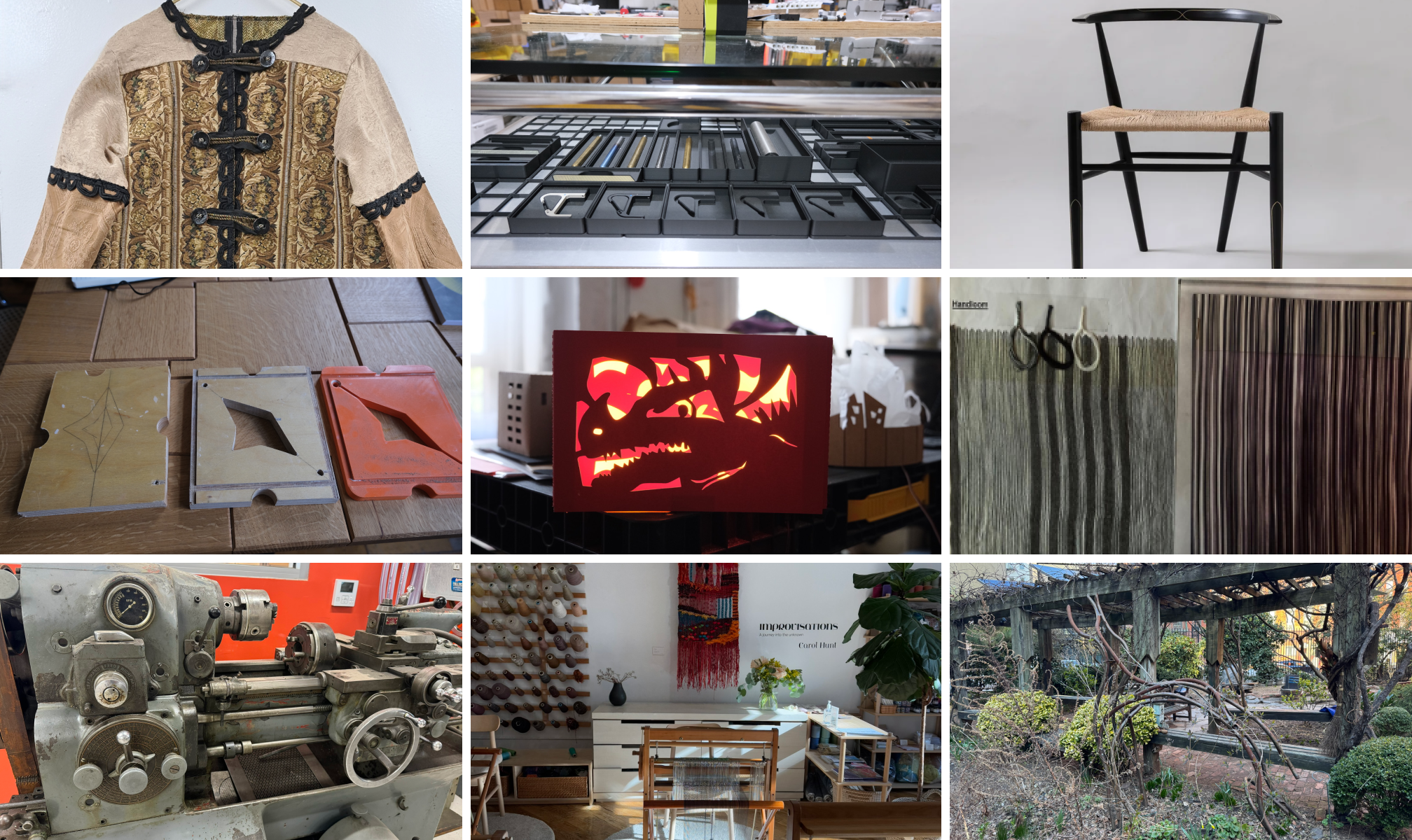}
    \caption{Select artifacts shared by expert participants including in-progress works, final pieces, craft tools, etc. Credits from top-left to bottom-right: Ann Kronenberg, Che-Wei Wang with Taylor Levy, Charlie Ryland, Gregory Beson, Jody Culkin, Sally J. Jones, Scott Van Campen, Yukako Satone with Carol Hunt, and Kathy Creutzburg with Mirabai Kwan Yin and Natalia Lesniak.}
    \label{fig:participants}
    \Description{These nine images from the study participants characterize the broad range of materials and techniques they use for their practices. Image credits belong to Ann Kronenberg, Che-Wei Wang with Taylor Levy, Charlie Ryland, Gregory Beson, Jody Culkin, Sally J. Jones, Scott Van Campen, Yukako Satone with Carol Hunt, and Kathy Creutzburg with Mirabai Kwan Yin and Natalia Lesniak.}
\end{figure*}

Participants were recruited using snowball sampling. We define an ``expert'' as someone with over five years of professional experience in their craft practice.
We sought to include a broad range of craft practices by targeting woodworking, fiber arts, and metalworking.
Their fabrication workflows enabled us to capture diverse tool-use and material engagements.
Table~\ref{tab:participants} summarizes our participants' years of experience, professions, primary craft techniques, and commonly used materials. 
Figure~\ref{fig:participants} provides some examples of their tools, in-progress works, and final pieces.

\subsection{Positionality Statement}

We acknowledge how our positioning as researchers in the United States (US) shapes the study design and findings. The research team represents diverse backgrounds across computer science, anthropology, and science and technology studies (STS). The first author maintains an amateur practice in machine knitting and woodworking, while all team members engage with digital fabrication and craft practices through educational and research-based contexts. This combination of hands-on craft practice, theoretical grounding, and technical experience informed both our participant recruitment and interview design, shaping questions that attended to computational tool design and material engagement. By snowball sampling from our network, we recruited participants predominantly located in urban and suburban areas of the northeastern US (N=21), though several (N=8) shared experiences living and practicing craft outside the US. All participants have access to shared workshops, university facilities, or private studios. While we aimed to recruit participants across diverse craft communities, our snowball sampling approach naturally drew from existing interconnected networks that may share perspectives and approaches to craft practices.

\subsection{Coding and Analysis}

The first author recorded all the interviews using video and audio for Zoom calls and just audio for in-person sessions.
Transcripts with diarization~\cite{radford_robust_2023} were generated locally using Whisper large-v2, then manually reviewed and corrected by the first author.

We developed a codebook by applying inductive techniques to our interview data.
The first three authors independently analyzed an initial set of three interviews (P2, P5, P7).
These three participants were selected to capture diverse perspectives, engaged materials, and professional backgrounds (e.g., independent artist, product designer).
Through this phase, we identified emergent themes and collaboratively developed low-level codes.

Following this, we used deductive coding by applying our initial codebook to the remaining 20 interviews as a method of ensuring that our codes were robust and repeated throughout the interviews.
The codebook was treated as a living document throughout the collaborative coding process; each interview was coded by at least two authors to ensure multiple perspectives were brought to the data and to build a shared context across the research team. 
The research team had six 2-hour meetings throughout the coding process to discuss new codes that emerged from the interviews and ideas for iterating on the initial themes.

Throughout the coding process, the first author grouped the low-level codes into higher-level code groups.
For example, our low-level code of \emph{productive accidents} related to discussions around accidents that are embraced as valuable parts of the workflow fit under the code group of \emph{improvisational workflows}.
Similarly, \emph{re-purposing materials} was placed under \emph{environmental sustainability} and \emph{influences from politics} under \emph{society influence}.
These code groups served as a foundation for our final thematic discussions and were iterated upon by the authors during the paper writing process (sample codebook in Supplementary A2).
\newcommand{\med}{Abstraction}
\newcommand{\nav}{Engagement}

\section{Findings: Orders of Convivial Relations}

\begin{figure}[ht!]
    \centering
    \includegraphics[width=\linewidth]{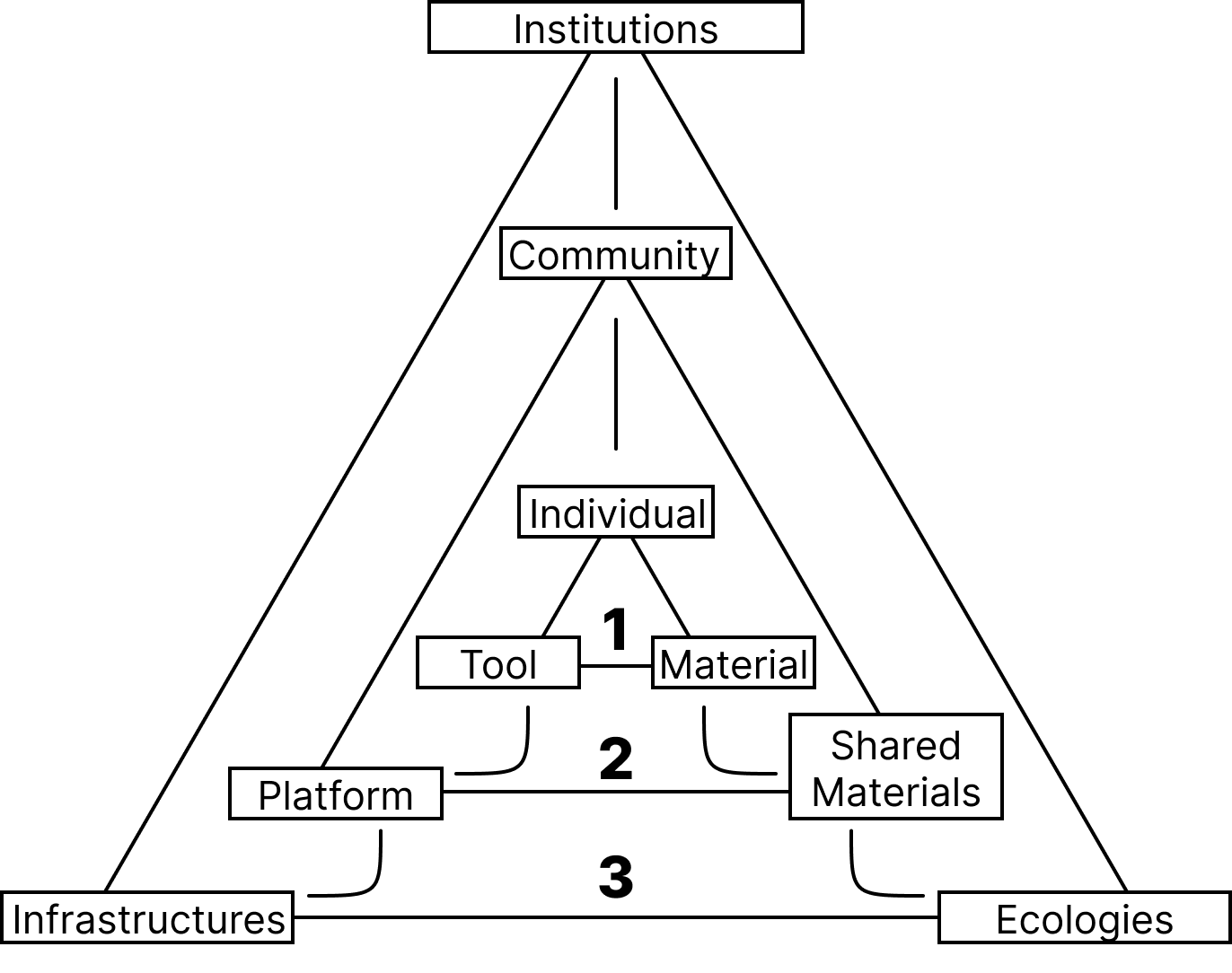}
    \caption{Our interviews reveal three interconnected orders of convivial relations in craft practices.}
    \label{fig:orders}
    \Description{This figure shows our 3 orders of convivial relations through concentric triangles (immediate, mid-range, extended).}
\end{figure}

In this section, we draw on empirical evidence from expert participants to analyze the orders of competing demands across immediate, mid-range, and extended relations in convivial craft practices. In Figure~\ref{fig:orders}, we illustrate the complexity of interactions across orders of convivial relations.
For each order, we examine how computational tools mediate craft relationships and then how craftspeople preserve, extend, and rework their material values and practices when their tools fail to operate in conjunction with their values. 

Participants describe the friction between material relations and values embedded into computational tools for fabrication.
Computational tools may enhance both speed and precision but they also simultaneously imposing rigid workflows.
Computational tools also limit direct interaction, fragment collective knowledge, and abstract away material life-cycles.
Craft workflows are characterized by a perpetual dialogue between craftspeople enacting their values while balancing between the efficiencies and constraints of computational tools.
Participants share their ongoing struggles to reconcile emergent material agencies with procedural linearity in tool-driven workflows.

\subsection{First Order: Immediate Relations} \label{sec:first_order}

Tools mediate the immediate relationship between craftsperson and materials---the hammer and chisel stand between sculptor and stone. Tools translate sculptor intent while conveying the stone's gradients of hardness, its lines of yielding and resistance.
% The relationship between sculptor and material remains intimate and immediate, grounded in physical co-presence and sensory feedback.

Whereas hand tools are extensions of the body, computational tools introduce abstractions that alter and attenuate relations.
As described in the hylomorphic model, digital tools assume that craft is a unidirectional transmission of preconceived plans onto materials~\cite{ingold_textility_2010}. This mentality is exemplified by the Star Trek replicator fantasy projected onto 3D printing, where objects materialize unbound by material history.
In contrast, our interviews reveal craft practices emerge from responsive partnerships with an active material world that ``pushes back'' on the design of the craftsperson.
\citet{dourish1999embodied} described how the computer mouse switches discretely between Heidegger’s \textit{ready-to-hand} and \textit{present-at-hand}. Craft tools, in contrast, move fluidly across these states, functioning simultaneously as shapers of material and \textit{sensors} that make the craftsperson aware of both tool and material.

Tools that preserve material dialogue shape both the craftsperson's workflow and the resultant artifacts.
This section examines how computational tools abstract material dialogue and how expert craftspeople navigate these limitations.

\subsubsection{\med{}} \label{sec:first_con}

\begin{quote}
    \emph{``[The computer] separates you from the material''} (P1).
\end{quote}

P1, a sculptor and fiber artist rooted in direct material relationships, illustrates the interposition of computational tools.
Her journey with computational tools began with techno-optimism born from early experiences with digital painting; these tools mirrored actions in the real world, and she noted that \emph{``moving a mouse around is very similar to moving a brush around.''}
However, this optimism faded when applying computational tools to sculpture and textiles.
Digital analogies break down when they fail to account for the tactile relations essential to her work:

\begin{quote}
    \emph{``I have a dialogue with the materials. Craft is about manipulating materials and the computer gets in the way''} (P1). 
\end{quote}

Digital abstraction decouples representation from reality, distorting proportionality and producing flawed assumptions about how a design will manifest. 
P6, a woodworker and educator, discusses challenges with scale when teaching students to translate 3D models into functional furniture:

\begin{quote}
    \emph{``[My students] get lost in 3D space and lose perspective of what it's going to be in the real world. For example, I teach a chair class, and often they'll say, `I want to make the seat height 20 inches....' It's in between a dining chair and a stool and your legs are hanging off''} (P6).
\end{quote}

Computer mediation changes individual style and preferences by pushing craftspeople towards workflows that align with software's affordances rather than material behaviors.
P1 notes that computers excel in a \emph{``flat [world of] color and form rather than materials,''} making computational tools compatible with 2D crafts such as machine embroidery.
Working on a 2D screen, however, is fundamentally at odds with her sculptural practice, as evident in the contrast between fabric draping and flat pattern design:

\begin{quote}
    \emph{``One of the primary ways that people design clothing is they take the fabric and drape it on the dummy. Flat pattern design is an industrial process.... For the most part, designers either draw or use dummies, and then somebody called a pattern maker comes in and does the flat pattern''} (P1).
\end{quote}

Relying on a tool’s features can stifle the serendipity that defines many craft practices.
P3, a product designer, notes:

\begin{quote}
    \emph{``[If craftspeople] spend all their time on the computer to understand how the material behave[s] and generate something based on that, [they] will kill the potential or the possibility to discover something different''} (P3).
\end{quote}

Finally, computational tools alter the craftsperson's relations to rhythm and time.
While P1 recalls disliking fiber arts as a child because they were \emph{``so time-consuming,''} many craftspeople find value in slower interactions.
P13, a weaver, frames this industrial drive for rapid fabrication as a cause for the loss of tacit knowledge, as \emph{``people forget how to do [weaving] by hand.''}
She describes teaching students to follow a natural, comfortable pace for hand-weaving in contrast to an accelerated workflow:

\begin{quote}
    \emph{``I ask [my students] to... start [with a] walking rhythm. Not like, `I'm going to coordinate [with] the machine.' That's too much. Just walking. And then the walking rhythm is very natural [for people]''} (P13).
\end{quote}

For P13, returning to a hands-on process is \emph{``very important [to] awake [humanity's] original mind.''}
P13 highlights how the mediation of time by machines is a fundamental shift that can distance craftspeople from an essential, deliberate way of being and creating with materials.

Across these examples, computational tools mediate immediate relations by creating distance between craftspeople and materials. Whether through abstracted away materials or computing-driven workflows, these tools prioritize predictability over dialogue and engagement. The result is not only a loss of creative opportunities but a fundamental reorientation of craft from material collaboration to predetermined execution.

\subsubsection{\nav{}} \label{sec:first_pro}

\begin{quote}
    \emph{``Listening to the wood when you're cutting it''} (P6).
\end{quote}

Craftspeople cultivate sensory and embodied engagement with their materials which helps them navigate the limits of computational tools.
For P6, bridging this gap requires listening to tactile and auditory cues that are nonexistent in computational tools.
Material sense-making is a learned skill responding to the material's feedback during craft:

\begin{quote}
    \emph{``You can listen with your body and get a lot from [the wood]. And you can hear it on a table saw, if you're cutting it the wrong way... you can hear the pitch change.... The wood's just like, `No' and then you'll look at it [and] it's chipping.... You flip it around, now you're moving with it and everything is cutting nicer and the sound is better''} (P6).
\end{quote}

Intentionality fosters flexibility. 
Craftspeople adjust to their material without rigid, preconceived plans.
P9, a jewelry designer, emphasizes a process of discovery by describing how designs emerge from material inquiry:

\begin{quote}
    \emph{``I have somewhere between 50 to 70\% of an image in my head knowing what I'm doing.... It is about the interaction that I have with the material itself.... Most of the time, the work comes out as I'm playing around with the material''} (P9).
\end{quote}

Improvisation enables critical discovery, refinement, and iteration before determining the final form.
P17, a woodworker who typically works with un-processed lumber, knows that plans are provisional until he can read the history written in the wood's grain, twists, and knots.
He shares that even \emph{``best laid plans going into a log can change real fast,''} as the final form emerges in negotiation with the material.

Adaptive workflows grant the craftsperson intentionality in their process.
P6 describes being in a \emph{``flow state''} where improvisation is paramount: \emph{``If you get too lost in a tool being broken or a material not working.... I usually just pivot really quickly.''}
Resilience demands \emph{``multiple ways to get to an idea''} (P6) in contrast to linear workflows with a single point of failure.
Similarly, P22, a mending artist and teacher, noted a shift in material engagements. While traditional techniques embrace flexibility, contemporary making processes stem from fixed expectations:

\begin{quote}
    \emph{``People think there's a right way of doing things. [But] I think the material will dictate what happens. If you have a needle, you can put it through silk and there's a lot of experiential learning involved''} (P22). 
\end{quote}

Making is material-led and unfolds through interactions that cannot be fully planned or anticipated in advance.
All making is tied to a particular medium and technique. P23 described how her \emph{``visual voice''} emerges across different modes of making:

\begin{quote}
    \emph{``In drawing, it's very clear to me.... I understand what my hand looks like when I draw it and it's very clear. Then, I go to something like ceramics, and I have to figure out what my hand looks like. When I say my hand I mean, `What's my visual voice in this new medium?' Your brain then has to translate your vision through your physical body, and there's a massive learning curve. I'm much quicker now when I pick up a medium. I can just find it much quicker, but it [still] takes material investigation''} (P23).
\end{quote}

A critical element of craft practice, along with much of its mystery, intensity, and allure, lies in this co-creative dialogue. P3 describes this as \emph{``quite an intense love story with this material where you hate and love [it] at the same time.''} 
These practices reveal how craftspeople maintain material dialogue despite computational mediation. By cultivating responsive workflows, they preserve the relationships with materials that define their practice. The material remains a central participant in the fabrication workflow and craftspeople develop a kind of tacit knowledge that enables them to respond to its changing demands.

\subsection{Second Order: Mid-Range Relations} \label{sec:second_order}

A second order of relations expands beyond an individual, concerning how computational tools mediate relations within craft \textit{communities} and the materials they share.
As craft studies and social anthropology have demonstrated (\textit{inter alia}, \cite{lave2001legitimate, lave_learning_2019, gordon_textiles_2011}), material knowledge and skilled practices are transmitted within communities of practice through guilds, apprenticeships, and workshops---through tactile and collocated forms of collaboration and training.
Computational tools alter relational networks, reconfiguring collaboration and communication through platforms and shared digital artifacts. Hands-on training is supplanted by online patterns and YouTube tutorials.
While this supports new forms of craft knowledge sharing, computational mediation transforms knowledge from a dynamic, social process into a static object.
Like many social platforms, interactions become a commodity owned by actors outside of the community. Platforms often incentivize over-productivity, performative making, spectacle, and gamification of craft practices.
Participants revealed that the most meaningful learning and collaboration happen through storytelling, demonstration, and feedback.
Tools that support rich, social, contextual knowledge transmission shape how craft legacy is preserved and enriched.
We examine how computational platforms can alternatively fragment and undermine craft pedagogy and collaboration, and how craftspeople work nevertheless to build and maintain effective networks and communities in the presence of computational tools and platforms.

\subsubsection{\med{}} \label{sec:second_con}

\begin{quote}
    \emph{``[My students] will get led by [the tool].... [There was] a new feature in [Adobe] Photoshop that you could bevel something. And overnight, everything was beveled.... For every client, whatever you're working on, you have to have bevels'' (P10).}
\end{quote}

Platforms alter how communities engage with shared materials by standardizing both techniques and aesthetic preferences. 
P10, a sculpture artist and professor, illustrates how platforms such as Adobe can reshape entire communities' relationship with their materials, not through collaborative evolution of a shared practice but through the availability of platform-specific features that can quickly become community-wide norms.
P10's recollection of the \emph{``era of the bevel''} demonstrates how a single platform feature can redefine a craft community's material language.

This phenomenon extends beyond design into craft education.
P8, a knitting studio manager, critiques how popular crochet kits like Woobles mediate the community's relationship with yarn by reducing or rendering invisible more challenging components of the workflow.
By making crocheting ``accessible,'' kits fragment the community's shared understanding of how to work with yarn, replacing personalized transmission of material knowledge with standardized instructions and inputs that fail to capture variations and personal techniques:

\begin{quote}
    \emph{``[Woobles] even do[es] that magic ring for you.... There are tips and tricks that work for different people, and you just have to figure out what that person needs. Whether it's `you need to twist your wrist in a certain way' or `you [need to] wrap it a different way' or `hold the tension a different way'''} (P8).
\end{quote}

Platforms are unable to effectively transfer tacit knowledge within a community.
P18, a crochet artist, articulates how platforms fail to transmit an embodied understanding of materials:

\begin{quote}
    \emph{``It's very hard to teach dexterity. It's very hard to teach how you hold your hands.... I think that's a problem with a lot of YouTube videos is that they don't really show you how to hold the yarn and how to hold the hook and how to get the tension right''} (P18).
\end{quote}

Platform mediation also affects how communities share critical knowledge about material safety, waste, and behaviors.
P4, who teaches digital fabrication, explains how platforms cannot adequately convey material risks:

\begin{quote}
    \emph{``So if it's something like a CNC machine where there are real consequences where there is an actual version of harm that can come, I would rather [it] be written and just emphasize `go as slow as you need to....' This is an important thing for safety''} (P4).
\end{quote}

Platform norms and self guided online learning contrast with collaborative safety culture that craft communities develop.
P12, a metalworker, further articulates the collective responsibility that emerges from shared relationships with materials: 

\begin{quote}
    \emph{``I've got hundreds of thousands of pounds of metal hanging over people's heads 24 hours a day.... That's the responsibility. And, you know, welders do take that seriously''} (P12).
\end{quote}

Computational platforms, while promising to connect communities, also mediate their relationships in ways that standardize techniques, fragment tacit knowledge, and erode archival practices that have been developed through generations of shared material engagement and knowledge sharing.

\subsubsection{\nav{}} \label{sec:second_pro}

\begin{quote}
    \emph{``It's not only a relationship with other people but it's a relationship with the space and also your relationship to yourself.... Crochet is just a means for the real magic that you know coming together brings''} (P16).
\end{quote}

Craftspeople also use computational platforms to sustain their communities and facilitate peer-to-peer knowledge sharing.
P16 described how crocheting functions as ``a means'' to cultivate a community of fiber artists who gather weekly at a coffee shop to engage with shared craft materials such as yarn, patterns, and techniques.
Regular social interaction coordinated through Discord enables participants to maintain relationships, both with each other and shared materials.
Platforms are not a substitute for material co-engagement, but rather an organizing mechanism to sustain the \emph{``real magic''} (P16) of collective material practice when craftspeople come together.
P2 similarly described how computational platforms enable connection with broader communities of craftspeople and hobbyists who provide feedback on his products and share their design modifications:

\begin{quote}
    \emph{``We started a Discord with just a few people and it's really nice seeing people share their projects because people make our stuff.... It's really interesting to see how people interpret our work''} (P2).
\end{quote}

These accounts show how craftspeople work around computational platforms to maintain hands-on material knowledge within remote communities.
Rather than using platforms as repositories of abstract knowledge, they are used to facilitate ongoing material dialogues that build collective understanding through exchange.
P19, a sculpture artist, described how materials serve as a common ground where distributed expertise converges onto a final piece:

\begin{quote}
    \emph{``If [people] see that you have a sketch, then you have an idea on paper. They know that there's something starting. We do collective brainstorming and sketching. Like any design process, even in computer programming, you go back and forth, revisit, redraw, change an area as the material changes or as the idea gets a little bit more refined or you understand your limitations''} (P19).
\end{quote}

Through exchange, individual material knowledge becomes collective.
Materials circulate within a community, accumulating expertise; each interaction adds to a shared vocabulary of possibilities and constraints.
Craftspeople build collective archives of material knowledge, sharing insights about materials that others incorporate into their practices.
P16's described their yarn library in this way:

\begin{quote}
    \emph{``[The yarn library] is the same concept as thrifting.... It's stocked by the community and anybody in the community can take [from] it.... That's my favorite place to get yarn if I can find something that works well''} (P16).
\end{quote}

Libraries do more than distribute materials; they preserve and transmit material histories.
P22, a mending artist who works primarily with donated scrap materials, makes these narratives visible:

\begin{quote}
    \emph{``I usually work on donations and scroungy, scrappy things to show that you can use materials and then embed those stories into a new thing and mend things visually, like loud, so that there is a statement''} (P22).
\end{quote}

These accounts demonstrate how craftspeople use platforms as organizing mechanisms to sustain collective practices. Shared materials hold collective memory, countering computational platforms' tendency to standardize them and strip away their histories.

\subsection{Third Order: Extended Relations} \label{sec:third_order}

Our third and final order of relations concerns how infrastructures (including those driven by computing) mediate the relation between institutions and the material ecologies they inhabit.
By institutions, we mean the extended networks and organizational structures that shape craft practice beyond individual craftspeople and their immediate communities, including markets and educational programs. These institutions rely on infrastructures such as global supply chains and distribution systems to create standardized, scalable, and economically efficient systems of production.

However, in doing so, they often abstract away the rich historical, cultural, and even political contexts of materials, treating them as interchangeable commodities isolated from the rest of the world.
Our interviews reveal insights into how expert craftspeople think about and engage with the these contexts of their materials and how their practices are shaped by what these infrastructures make visible (or invisible).
This section first examines how infrastructures obscure these relationships and then how craftspeople actively build connections between their practice and their material's ecologies.

\subsubsection{\med{}} \label{sec:third_con}

\begin{quote}
    \emph{``Craft is just labor glorified. For whatever reason, we decided that we need to put a pretty hat on a certain set of labor''} (P17).
\end{quote}

The infrastructures that connect craft to its material sources mediate the relationship between institutions and material ecologies, often at the expense of the material world.
P17, a woodworker, shares how supply chains hide the labor necessary to produce craft materials, limit access to environmentally sustainable resources, and standardize cultural norms.
P17's specializes in green woodworking, or \emph{``working with freshly felled lumber as opposed to kiln-dried store-bought stuff''} (P17).
Working in this way creates a direct and intimate relationship with the material world that infrastructures such as supply chains obscure.

Computational infrastructures can sever the labor that produced craft materials, enacting forms of erasure and violence against the communities that---in producing these materials---are enacting their own forms of craft.
P17 argues that educational infrastructures erase the contributions of people of color in craft history:

\begin{quote}
    \emph{``[Craft] is entirely dominated by a bunch of white men. And historically, we're told that that's exactly who made everything that is in this country.... [Yet] estimates show that around the time of the Civil War in the South, 80\% of skilled labor in the United States was done by Black people.... It's not even close''} (P17).
\end{quote}

Institutional erasure simplifies material histories that ignore the complex and often violent ecologies from which craft objects emerge.
P1, a fiber artist, shares P17's perspective by describing how the feminist art movement emerged as a direct response to infrastructures devaluing and erasing the craft practices historically associated with women's labor:

\begin{quote}
    \emph{``95\% of the people teaching at the college level were men and 95\% of the people taking classes were women and there was a real divide... this spawned the feminist art movement.... One of the things that [the movement was] talking about was using women's craft processes in your art which are mainly fiber processes''} (P1).
\end{quote}

This pattern of infrastructural abstraction extends beyond labor to environmental and social costs in global supply chains.
P5, a furniture designer, critiques the extractivism of the global food market, where materials are stripped of their local context to serve others around the world:

\begin{quote}
    \emph{``What can I do to give farmers more money out of their heirloom corn [without] falling into the trap of the gourmet restaurant version of `We're going to take it to... all the Michelin star restaurants?' And that's extractivism. That's taking their good grain way from the communities to fancy restaurants''} (P5).
\end{quote}

This process renders the material's origin and journey invisible to the end consumer.
P16, a fiber artist, experiences this on the consumer side, expressing frustration with the illusion of ``sustainable'' choice when sourcing yarn for textile projects:

\begin{quote}
    \emph{``I know that [with] cotton, there's this weird Catch-22 with sustainability because cotton uses a lot of water to grow, but it's also in places that are famous for droughts. California produces so much of our cotton in the U.S. and literally, they have to take short showers over there.... If you want to be sustainable, use natural fibers, but then there are stories behind the natural fibers every single time.... No ethical consumption, no matter what''} (P16).
\end{quote}

The global market is designed to make complex material histories invisible. The craftsperson is presented with a material that is deceptively neutral and context free.
Price-driven markets create infrastructures that crowd out craftspeople's access to less-processed materials.
P17 shares how sourcing a wooden log from the forest compares to the convenience of the industrial alternative:

\begin{quote}
    \emph{``I'm very curious... [to] find ways that you can work with kiln-dried lumber which is a little bit more easy to find, easy to kind of understand. You can buy [it] in smaller quantities. It's just a little bit more user-friendly in a certain sense. But I want the same material understanding and relationship you get from working with a tree when you're doing that''} (P17).
\end{quote}

Similarly, P11 describes how regulatory frameworks designed for industrial production make recycling bureaucratically (note, not technically) challenging:

\begin{quote}
    \emph{``Let's design a machine that can... take all this scrap and actually put it back into a yarn that is usable.... We can do a better job of actually selling it to the hand weaver and the crafter too, than putting it back into a fabric. The problem with... selling [recycled fibers] globally is that you have to have every single item that is in the fabric noted in the bill of lading. If it's got cotton, acrylic, polyester, nylon, you've got to put the amount in it too. So it's very difficult selling any recycled fabrics across borders... [governments] charge tax on these things. It's got nothing to do with the people; it's got to do with how countries make their money''} (P11).
\end{quote}

Material costs are not produced via the invisible hand of the market. There are concrete political and economic decisions that shape prices.
As P5 notes, synthetic materials are artificially cheap, propped up by government subsidies that natural materials lack:

\begin{quote}
    \emph{``When we think about plastics, we really are talking about the most subsidized material in the world. We rarely pay the real value of a piece of plastic. How in the world is a plastic that is like digging up petrified fossils several hundred years where the earth or in the ocean be cheaper than [natural materials like corn]? It's like we've created a whole system where whole governments subsidize that extraction''} (P5).
\end{quote}

Infrastructural mediation reduces knowledge tied to regional ecologies, replacing local traditions and practices with a standardized, globalized model of production.
P21, a furniture maker, reflects on how material availability historically shaped regional craft in ways that are now lost with easy access to standardized materials:

\begin{quote}
    \emph{``There's something to say about the fact that all the furniture from this region was made out of pine and all the furniture from this region was made out of oak. That's not to say maybe those were the best species. That's to say, well, that's all they had there''} (P21).
\end{quote}

Modern supply chains makes place-based knowledge less relevant, favoring a universal approach where any material can be sourced for any project.
The result risks reducing distinct craft practices to the execution of universal patterns, rather than a dialogue enriched by the situatedness of the craftsperson.
P1, a fiber artist, identifies this as a distinct trait of the American market, which historically favors repeatability and de-skilling over in-situ dialogue with materials:

\begin{quote}
    \emph{``The tradition in the United States was you got directions, you got a kit, and you made what somebody else designed. And the artware movement emphasized improvisational knitting, crochet, and embroidery''} (P1).
\end{quote}

These accounts reveal how infrastructures from educational institutions to global supply chains mediate the relationships between markets, craftspeople, and their material worlds.
By erasing complex labor histories and environmental costs, standardizing cultural practices, and creating economic barriers to materials, these systems present a deceptively neutral and standardized material reality.
This constrains the possibilities for a conversational engagement with materials, one that challenges the very autonomy that lies at the heart of many craft practices.

\subsubsection{\nav{}} \label{sec:third_pro}

\begin{quote}
    \emph{``We have a whole financial structure to start paying [farmers].... I don't see this as charity. This is part of a business plan'' (P5).}
\end{quote}

In response to infrastructural mediation, many of the expert craftspeople we interviewed shared how they actively shape their practices to (re-)establish and celebrate the deep connections between their work, communities, and material ecologies.
They navigate the pressures of the market by creating alternative systems of value and incentives that retain the histories and relationships typically made invisible by infrastructures.
The work of furniture designer, P5, serves as a powerful model for this approach.
Raised with a deep appreciation for agriculture, farming communities, and material reuse, his practice intervenes through craft to drive ecological restoration and economic justice in an indigenous community near where he grew up.

One approach to navigate infrastructures mediating between institutions and material ecologies is to transform a material's history and origins into a source of value:

\begin{quote}
    \emph{
    ``There's this concept in wine-making called `the terroir,' which translates loosely to the territory... the geographical, geological, climate, and cultural conditions that create a certain kind of crop.... What is the terroir of plastics? It has nothing, it's insipid.... We don't do that with [synthetic] materials, [but] we're always asking that with food in restaurants'' (P5).
    }
\end{quote}

By centering the unique story of materials such as heirloom corn husks and agave fibers in his own practice, P5 describes a new kind of market based on authenticity and connection rather than perfection.
The same principle of valuing an object's lived history is central to P22, a mending artist who works with donated and used textiles.
For her, ``flaws'' are not defects but a critical and valuable part of a material's properties that cannot be ignored:

\begin{quote}
    \emph{``I'm wearing a really silly sweater. And I wear that a lot because when I broke my arm in the pandemic... that was the only sweater that I could wear. I sliced [it] up here [points to her shoulder] and then I sewed it back together.... Whatever value we attach emotionally to how we want to be perceived, you always tell a story with what you wear''} (P22).
\end{quote}

Making material histories visible is straight-forward when consciously operating locally, scaling down production to engage deliberately with material ecologies.
P5's economic model, for example, is built around the specific needs and resources of a single agricultural community.
He actively resists the market pressure to scale infinitely, recognizing that the integrity of his work is tied to its specific context and location:

\begin{quote}
    \emph{``The Minister of Agriculture of India came to us and [asked], `How do we bring this to India? How do we roll it out? How can we scale this up, industrialize this....' I was like, `You can't, and you shouldn't....' Scalability [works] to a certain point, but not too much because you start to fall in the trap of an industrial production if you go too big''} (P5).
\end{quote}

The commitment to place and locality extends to other participants as well, who find creative potential in the overlooked or ecologically disruptive materials within their immediate environment.
P19, a sculptor and community garden coordinator, describes using an invasive plant species as both a material and form of ecological intervention:

\begin{quote}
    \emph{``We're working on a new sculpture in bamboo and it's the first time... trying to use an invasive species, an invasive plant in the work.... One of the community gardens here has bamboo that's kind of taken over and it's right along the facade of the building, so it needs to go''} (P19).
\end{quote}

This practice transforms an ecological constraint into a creative resource, grounding the craft practice in the reality of its environmental contexts.
Similarly, P21, a furniture maker, describes how the limiting circumstances of the COVID-19 pandemic forced him to abandon industrial supply chains and engage with his urban ecology in a new way:

\begin{quote}
    \emph{``When shutdown happened because of COVID, I was in a little Boston apartment, and I didn't have any tools, and I didn't have access to a shop. All I did was I walked into a park nearby and I grabbed limbs that had fallen out of trees and I learned how to carve spoons''} (P21).
\end{quote}

Another way craftspeople navigate digital mediation is by prioritizing legacy and creating durable artifacts that resist short-term market pressures.
P5 shares the story of Silvestre (pseudonym for anonymity), a community elder who spearheaded reforestation with no expectation of living long enough to see its completion and impacts on the community:

\begin{quote}
    \emph{``He was in his 70s, digging holes along these mountains, planting agaves... to revert the erosion to maybe one day be able to grow their corn again.... This is work that [Silvestre] will never see. And so this... selflessness of actions... is what's really powerful here''} (P5).
\end{quote}

Beyond ecological restoration, craftspeople create self-sustaining organizations that embed values of upcycling and creative reuse into broader craft communities.
P23, a textile artist, builds legacy through a collective she founded:

\begin{quote}
    \emph{``I made a little collective in Berlin with some of my friends.... We would collect scraps from people... and we would do these open free textile upcycling workshop days where we brought all of our sewing machines... and we would invite people in off the street.... It was all about fixing the stuff you've got, making new things, creating whatever you want out of nothing''} (P23).
\end{quote} 

These practices reveal how craftspeople actively resist distancing and abstraction by centering material histories in their practices. Where global supply chains present materials as neutral commodities, craftspeople transform origins and lived histories into sources of value and inspiration. Where markets create pressures towards infinite scaling---a language that computing discourse and practice all too often uncritically adopts and reproduces---craftspeople deliberately limit production to maintain ties to local communities and ecologies. Where economic systems tip the scale in favor of industrial materials, craftspeople share and leverage the creative potentials of local and overlooked resources. Where these commitments to locality, engagement, and accountability are often out of step with both the affordances and assumptions of computational thinking, these differences introduce an intractable (yet perhaps addressable) tension into the relations between craft practices and computational tools and infrastructures.

\section{Discussion}

It is important to note that the orders above are descriptive devices intended to name categories and scales of relations often missing from tool-centered accounts. They are not rigid constructs, and the craftspeople in our study frequently and skillfully navigated between them.
We have seen several examples where the craft method itself is tied to a specific extended relation. In choosing to work with only bamboo from the community garden, P19 is articulating a commitment to locality---and pointedly, a choice \textit{against} other kinds of relations.
In committing to certain kinds of sourcing relations---say, wool from a particular farm, whether around the corner or around the world, knitters make `local' choices that respond to and intervene in wider systems and infrastructures. 
Fluidity is a hallmark of a holistic craft practice in which actions in one order shape and are shaped by the other orders.
Working in pluralistic ways with and from materials may require more pluralistic ways of working with and from \textit{worlds}, and the wider ecologies, infrastructures, and institutions that constitute them.
This reach is essential to both craft practice and the forms of value that accomplished objects of craft (e.g., the table, the knitted garment) hold.

Participants in our study repeatedly shared how computational tools abstract away material responses essential to their engagement with craft.
They also described how design and communication platforms reshape knowledge sharing and material reuse within their communities, to mixed effects.
Finally, participants described how infrastructures such as global purchasing networks (but also `global' educational institutions, like our own) could detach materials from their histories and origins, abstracting away or rendering invisible relations essential to their practice.

But what might a \textit{different} world of computational tools and infrastructures look like?
Can we identify design principles that lead to more convivial relations in craft and potentially other fields? 
The sections that follow explore what a more convivial world of design---a world in which computational tools and infrastructures
deepen and enrich rather than abstract away and attenuate material relations---might look like.

\subsection{Materials as Collaborators}

Illich's concept of conviviality centers on tools that \emph{``[enlarge] the range of each person's competence... limited only by other individuals' claims''}~\cite[p.6]{illich_tools_1973}.
His framework focuses on preventing tools from controlling individual human action, advocating for the design of tools that enhance rather than diminish autonomy.
As careful engagement with craft practice and a subsequent strain of `more-than-human' thinking within and beyond HCI~\cite{wakkary_things_2021} suggests, however, this concept may need to be updated and expanded. As Ingold describes, materials assert their own agency and constrain human autonomy in ways that craftspeople learn to accommodate, listen to, and ``correspond'' with~\cite{ingold_correspondences_2021}. Expert craftspeople in our study described their practice in terms of ``collaboration'' or ``conversation'' with material worlds and ecologies, and they expressed forms of care, learning, and resonance with such worlds. This stands in some tension with the languages of autonomy and mastery still present (and we believe problematic) in Illich's work. If autonomy for Illich names a (partial) independence from industrial forms and scales of production, it should \textit{not} imply a separation from the material world writ large---or what amounts to the same thing, an `external' relationship of mastery, self-sufficiency, and control, in which the human actor (the only `agent' on the scene) holds and plays all the cards. This version of conviviality would indeed be fatal to craft of the kinds studied here, and could go some ways towards replicating the hylomorphic model of design criticized by Ingold.

Computational tools often abstract away human-material relationships, introducing both temporal and spatial distance between a craftsperson's decisions and the material's responses~\cite{twigg-smith_tools_2021, batra_composing_2025}.
With hand tools, there is an immediate feedback loop between a craftsperson's actions and the tool's interaction with the material that allows for real-time adjustments and decision-making.
In contrast, computational tools have longer feedback loops that perform actions on a material far removed from the material's responses.
Expert craftspeople from our interviews characterized this distance not as a loss of autonomy in Illich's sense, but as a constraint on their ability to listen to, learn from, and respond to material behaviors.
When computational tools demand predictable and ``perfect'' material behavior, they disrupt the very capacity for craftspeople to engage in the ongoing dialogue that defines skilled craft practices.
Designing convivial computational tools for fabrication therefore requires a fundamental rethinking of how computing can support (not replace through automation) this correspondence between craftspeople and materials.
We explore this challenge through three design principles:\\

\noindent \textbf{Design Principle 1. Embrace ``Imperfect'' Materials}:
P1's critique of computers excelling in 2D (rather than 3D) forms suggests an expansion of digital material representations and behaviors beyond standardized materials such as high-grade steel and wood.
For example, embracing digital material databases such as \citet{larsson_procedural_2022} champions material ``imperfections'' by scanning materials in their natural forms (for example, wood with knots and cracks) as part of the models.
Similarly, work from \citet{tamke2021treeto} explores digitizing a variety of wood grades for high-performance building products.
Rather than pursuing perfect and intractable material simulation (see \emph{Sim2Real} gap in robotic manipulation~\cite{peng2018sim}), this offers a practical path forward: make material ``imperfections'' legible and navigable to both craftspeople and computational tools. As shown in our interviews, craftspeople already interpret, for example, grain patterns and wood knots on their material as a part of their practice. Developing alternative computational abstractions for processing and representing such ``imperfections'' could enable novel computationally-supported craft workflows such as re-generating designs around a paper's coffee stain for inkjet printing or flagging low-durability areas of the material to support design placement for the CNC machine.\\

\noindent \textbf{Design Principle 2. Develop ``Fuzzy'' Models for Fabrication}:
P9's description of having only 50 percent of a design before fabrication suggests that computational tools could be made more attentive to the emergent material properties that fill these gaps.
Starting with ``fuzzy'' models for fabrication (could look like annotations, construction lines, or think-aloud utterances) allows craftspeople to specify their design intent not just as a shape but with higher-level goals such as the object's functional constraints and aesthetic qualities. Computational tools could then dynamically adjust tool-paths based on real-time material feedback, algorithms could propose design modifications when materials resist operations, and interfaces could visualize material preferences as constraints that guide design opportunities, all while maintaining the core design intent. \citet{read2024endgcode} and \citet{bourgault_millipath_2024} exemplify this principle by showing how computational tools can support design adjustments by directly (and continuously) engaging with materials such as PLA filament or clay.\\

\noindent \textbf{Design Principle 3. Synchronize Fabrication Pacing with Natural Rhythms}:
P13's emphasis on weaving following a \emph{``walking rhythm''} reveals how meaningful material engagement requires adapting fabrication pacing to match the natural rhythms of both humans and materials rather than \emph{``fostered addiction to speed''}~\cite{illich_tools_1973}.
By natural rhythm, we refer to the temporal dynamics through which materials respond (e.g., clay hanging due to gravity, yarns loosening after releasing tension) and at which craftspeople can perceive, interpret, and adjust for those responses in the design. Computational tools could support this dialogue by adjusting not just the fabrication speed but its pacing through pausing, sequencing, and emphasizing based on both material and human responses (similar to human-to-human conversation patterns~\cite{chui2005temporal}).
Existing work from \citet{subbaraman_playing_2024} and \citet{kim_machines_2017} describe how this adaptable pacing can create new possibilities and invite craftspeople to feel and respond to the material in real-time.

\subsection{Seeing (And Making) Within and Across Orders}

Our findings reveal how conviviality manifests distinctly within and across the different relational orders that constitute craft practices.

At the immediate order (Section~\ref{sec:first_order}), conviviality means maintaining direct and adaptive dialogue between craftsperson and material \emph{``through the use of tools that [an individual] actively masters, or by which he is passively acted upon''}~\cite{illich_tools_1973}. P6 describes this quite literally as \emph{``listening to the wood;''} when the material moves through the table saw, the craftsperson must adjust their technique based on auditory and tactile feedback. Computational tools often disrupt this dialogue by imposing predetermined workflows that treat materials as uniform and predictable. P3 warns that an over-reliance on such tools can \emph{``kill the potential or the possibility to discover something different.''} Rather than demanding materials conform to these tools, convivial tools would preserve the material's capacity to \emph{``push back''} by resisting and reshaping the craftsperson's design and fabrication workflow. This means supporting a kind of adaptivity where, as P9 describes, craft practices emerge through \emph{``interaction with the material itself.''} Convivial tools should therefore relay material feedback back to craftspeople, enabling their adjustments and dialogue with materials.

At the mid-range order (Section~\ref{sec:second_order}), conviviality emerges through the collective practices that preserve \emph{``creative discourse among individuals''}~\cite{illich_tools_1973}.
Organizing weekly crochet gatherings (P16) or sharing ideas with others (P2) through Discord are examples of how platforms can serve as organizing mechanisms for sharing the knowledge gained from immediate material engagements.
Discord is convivial in that the platform enhances individual autonomy through community-driven knowledge sharing that can be both evolving and personal to each craftsperson.
In contrast, crochet kits, providing materials, prepared magic circles, and uniform instructions, undermine community knowledge development by abstracting away and standardizing the nuances of the practice.
This suggests that convivial tools should encourage digital spaces where craftspeople can document and share their adaptive, contextual workflows rather than relying on mechanical instructions.

At the third order (Section~\ref{sec:third_order}), conviviality requires \emph{``the intercourse of persons with their environment''}~\cite{illich_tools_1973} by maintaining connections to the broader social and ecological relations within which materials are inherently embedded.
The craftspeople in our study shared how infrastructures such as the material supply chain can impose abstractions across the orders of relations.
P17 critiqued craft education for erasing labor histories, arguing that skilled labor has historically been dominated by people of color despite narratives crediting white craftspeople.
P16 described the illusion of sustainable material choices, explaining how seemingly natural fibers such as cotton mask environmental costs such as water usage in drought-prone regions, leading to their conclusion that ethical material sourcing is impossible.
Rather than treating materials as interchangeable consumables, convivial tools should create a relational trace that preserves and celebrates the extended relationships that give materials their form, behavior, and value.
This requires making visible (for example) the labor conditions, environmental impacts, and cultural practices of materials that computational tools often abstract away.
This contrasts with the industrial tools critiqued by Illich that create dependency on managers and hide the true environmental and social costs of production, making the infrastructures behind material extraction and sourcing invisible~\cite{star_ethnography_1999}.

This expanded understanding of conviviality challenges computational tools to facilitate not just human-tool relationships but the broader networks of the material world across its relations.
Design principles which support this goal include the following:\\

\noindent \textbf{Design Principle 4. Support Evolving Collective Knowledge}:
P16's Discord community demonstrates how convivial computational tools could facilitate craft knowledge sharing through non-prescriptive methods that preserve the situated nature of craft expertise.
Rather than simply sharing finished artifacts or standardized instructions, computational tools could be designed to capture and transmit the improvisations, adaptations, and personalized techniques that emerge during the fabrication workflow; the craftspeople in our study describe such elements as critical to their practices.
Ravelry~\cite{ravelry} exemplifies how craft knowledge can be collectively shared and iteratively built upon, reflecting the dynamic and nuanced nature of craft practices.
Similarly, \citet{kim_mosaic_2017}, \citet{batra_composing_2025}, and \citet{savage2025fedt} demonstrate that sharing works-in-progress enhances collaboration by making visible the evolving nature of design processes which informs expertise.
These examples illustrate how computational tools could go beyond just hosting content to enabling craftspeople to share their experiments, situated decisions, and responses to material behavior. Such features align with convivial principles by sharing diverse and idiosyncratic practices over prescriptive, one-size-fits-all documentation artifacts.\\

\noindent \textbf{Design Principle 5. Bridge Knowledge Across Orders}:
P5's description of the \emph{terroir} demonstrates how material knowledge can transfer across domains as immediate material properties (how corn husks behave when laser cutting), connect to community knowledge (harvesting techniques), and ecological contexts (soil conditions and farming practices).
To reflect this conversation across orders, convivial tools could adapt their behaviors based on feedback from all three orders simultaneously.
When material properties shift unexpectedly (first order), this could contribute to community knowledge about that material type (second order) while expanding understanding of how that material's source and distribution impact the unexpected material behavior (third order).
Drawing on examples such as consumer labels for resources~\cite{young_responsible_2018} or time-to-market metrics for semiconductors~\cite{ning_supply_2023}, computational tools for fabrication could similarly make visible such connections across orders of relations.
Such labels for materials could go beyond provenance to surface multi-order metadata, linking material behavior to its sourcing conditions, processing history, and community knowledge passed between craftspeople. Preserving these convivial relations could support craftspeople in seeing and interpreting the upstream factors that impact their workflows.\\
    
\noindent \textbf{Design Principle 6. Implement Fault Tolerance}:
P6's approach to having \emph{``multiple ways to get to an idea''} demonstrates a kind of fault tolerance that operates across the orders of relations. 
For example, when a router bit breaks during woodworking, a craftsperson may pivot to hand tools to maintain the project vision.
When community knowledge archives lack understanding about working with a particular wood species, the craftsperson may draw on their understanding of the tree's growth patterns to inform their approach---and perhaps feed back craft needs to local wood lots and timber operations.
Here, fault tolerance refers to the adaptability (not redundancy) of craftspeople and computational tools to shift strategies when a given method, material, or knowledge source becomes unavailable. Prior work on ``portability,'' as explored by \citet{roumen_portable_2023}, demonstrates how computational tools can support fault tolerance across different fabrication machines and contexts.
Convivial computational tools could extend ``portability'' to suggest alternative workflows, compatible materials, and parameter adjustments based on the diverse contexts drawn from immediate, mid-range, and extended relations.
This approach re-frames unexpected behavior in workflows not as failures but as opportunities to draw from and interpret knowledge from all the orders of relations.\\
     
\noindent \textbf{Design Principle 7. Prioritize Modularity with Shared Data}:
P23's ability to transfer techniques across different media from drawing to ceramics demonstrates how craftspeople develop modular approaches that can be adapted to new contexts.
Rather than creating monolithic applications that lock craftspeople in predetermined workflows, convivial computational tools could embrace modular architectures that allow integration across orders~\cite{feng_cameleon_2024}.
For example, a laser cutting tool could export material parameters (first) that integrate with community design libraries (second order) and material sourcing filters from supply chain databases (third order).
This modularity enables craftspeople to construct personalized workflows that maintain shared data between immediate material interactions, community knowledge sharing, and material ecologies.
This kind of compatibility calls for convivial computational tools to reflect the interconnectedness of craft practices, building on research from \citet{litt2025malleable} and \citet{mackay_interaction_2025} that demonstrate how shared data between tools strengthen (rather than constrain) creative workflows.\\

Beyond their utility in traversing across scales and orders, the design principles address questions of tool uptake and accessibility. Illich emphasized that \emph{``almost anybody can learn to use [convivial tools], and for his own purpose''}~\cite[p.64]{illich_tools_1973}, yet computational tools present significant barriers to entry. Three principles address this challenge: Principle 4 (collective knowledge) by documenting expert navigation around materials; Principle 6 (fault tolerance) by supporting learning by experimentation; and Principle 7 (modularity) by enabling the gradual integration of tools into existing workflows. Together, these principles pursue accessibility through transparency, foregrounding craft expertise and material relations.
\section{Conclusion}

In this work, we have explored how computational tools for fabrication can be re-designed for conviviality by better aligning with the long-standing needs, practices, and values of craft communities.
Illich understood that not every tool should operate this way:
\emph{``What is fundamental to a convivial society is not the total absence of manipulative institutions and addictive goods and services, but the balance between those tools which create the specific demands they are specialized to satisfy and those complementary, enabling tools which foster self-realization''}~\cite[p.32]{illich_tools_1973}). However, creating spaces for the design and development of convivial tools within HCI better aligns with the needs and practices of craft communities specifically, and inspires convivial tools in other domains of computing.

From this work, we learned how craftspeople navigate computational abstraction to preserve what we describe as a conversation between craftspeople, tools, and materials.
This dialogue emerges in distinct yet interconnected ways across the orders of relations from immediate engagements with materials to the extended engagements with their ecologies.
With this framing, we argue how an expanded understanding of conviviality for computational tool design can support the relations that are essential to craft practices and potentially other related domains such as digital fabrication pedagogy, where material's provenance and supply chains remain underrepresented in curricula despite shaping project outcomes.

Beyond craft, this work offers broader questions for computing and its own relations to the material world around it. How might computational tools abstract away or support engagement with computing \emph{materials}? How might platforms constrain or enable collective knowledge sharing around such materials? Lastly, how might infrastructures trace or fail to consider these materials ecological dependencies?
Consider software development: programming languages determine what material feedback (e.g., memory leaks, CPU cycles, energy usage) developers can perceive and respond to; application program interfaces shape how teams can adapt code for their own contexts and uses, obscuring interactions with hardware and processing; and the cheery interfaces of large language models abstract away the data centers, energy and water, and mineral supply chains required for their training and deployment.
Where recent work in sustainable and post-growth HCI examines the environmental and planetary impacts of computing~\cite{sharma2025sustainability, sharma2023post, soden2021we, liu2022towards, rodgers2023nature, laurell2025exploring, mandel2024designing}, conviviality may offer a complementary actor-based perspective that considers whether and how computational tools may deepen human engagement with the `stuff' of computing and the world---or alternatively abstract away these relations for exploitation and centralized control.
This question, central to Illich and other critics of industrial capitalism, remains fundamental to understanding (and critiquing) computing's role in the world today.
While we leave these questions for future work, we hope this paper opens pathways toward more convivial computational tools that support richer, more creative, and more accountable engagements between humans and the material worlds they inhabit.

\newcommand{\ulink}[2]{\href{#1}{\uline{#2}}}
\sloppy
\begin{acks}
We would like to thank our 23 participants for sharing their expertise and wisdom with us: \ulink{http://annekata.com/}{Kathrin Achenbach}, Paul Anderson, \ulink{https://gregorybeson.com/}{Gregory Beson}, Trevor Cross, \ulink{https://www.kathycreutzburg.com/}{Kathy Creutzburg}, \ulink{https://www.jodyculkin.com/}{Jody Culkin}, John Fletcher, \ulink{https://thecraftylounge.com/}{Lori Gaon}, \ulink{http://www.christopheguberan.ch}{Christophe Guberan}, \ulink{https://eric-hagan.com/}{Eric Hagan}, \ulink{https://pattyharris.cargo.site/}{Patty Harris}, \ulink{https://www.kolinjamesfurniture.com/}{Kolin James Schmidt}, Sally Jenkyn Jones, Ann Kronenberg, \ulink{https://www.fernandolaposse.com/}{Fernando Laposse}, \ulink{https://people.math.wisc.edu/~gemeyer/airsculpt/hyperbolic2.html}{Gabriele Meyer}, \ulink{https://www.instagram.com/magic_circle_bk/}{Rosie Reyes}, \ulink{https://www.jackiericcio.com/}{Jackie Riccio}, \ulink{https://www.instagram.com/charlie.ryland/}{Charlie Ryland}, \ulink{https://loopoftheloom.com/}{Yukako Satone}, \ulink{https://moeinedin.com/}{Moeineddin Shashaei}, \ulink{https://www.makerspace.nyc/}{Scott Van Campen}, and \ulink{https://cwandt.com/}{Che-Wei Wang}. We also thank Niti Parikh, Sebastian Bidegain, and the Craft@Large community for the early conversations and introductions and our labmates in the Computing on Earth Lab and Matter of Tech Lab for their feedback on the paper.
\end{acks}

\bibliographystyle{ACM-Reference-Format}
\bibliography{refs}

% \input{src/07_appendix}

\end{document}